\pdfoutput=1

\documentclass[11pt]{article}

\usepackage[preprint]{acl}

\usepackage{placeins}
\usepackage{pgfplots}
\pgfplotsset{compat=1.17}
\usetikzlibrary{arrows.meta, positioning, patterns, shapes.geometric}
\usepackage{float} 
\usepackage{booktabs}

\usepackage{times}
\usepackage{latexsym}
\usepackage{placeins}

\usepackage[T1]{fontenc}

\usepackage[utf8]{inputenc}

\usepackage{microtype}

\usepackage{inconsolata}

\usepackage{graphicx}

\title{"Can You See Me Think?" Grounding LLM Feedback in Keystrokes and Revision Patterns}

\author{
\textbf{Samra Zafar}\textsuperscript{\dag} \quad 
\textbf{Shifa Yousaf} \quad 
\textbf{Muhammad Shaheer Minhas} \\
National University of Computer and Emerging Sciences
}

\begin{document}
\maketitle
\def\thefootnote{\dag}\footnotetext{Corresponding author.}

\begin{abstract}
As large language models (LLMs) increasingly assist in evaluating student writing, researchers have begun questioning whether these models can be cognitively grounded, that is, whether they can attend not just to the final product, but to the process by which it was written. In this study, we explore how incorporating writing process data, specifically keylogs and time-stamped snapshots—affects the quality of LLM-generated feedback. We conduct an ablation study on 52 student essays comparing feedback generated with access to only the final essay (C1) and feedback that also incorporates keylogs and time-stamped snapshots (C2). While rubric scores changed minimally, C2 feedback demonstrated significantly improved structural evaluation and greater process-sensitive justification.
\end{abstract}

\section{Introduction}

Large Language Models (LLMs) are fast beginning to play a central role in writing instruction in all educational settings, providing scalable, accessible, and frequently informative answers to student essays. These have been adopted due to convenience and the rising evidence of the ability of AI- generated feedback to aid revisions, increase confidence in students, and improve writing \cite{meyer2025feedback,zafar2025cognition}. However, despite their promise, a major limitation persists: most LLM-based feedback systems focus exclusively on the final submitted draft, overlooking the complex, recursive nature of the writing process itself.
This fixed assessment model levels out the depth of the meta-cognitive work that goes into writing. There is nothing linear about writing, where there are many pauses and false starts, revisions, and hesitations and re-formulations \cite{lee2024cognitive}. Ignoring the acquisition of the final text, the LLMs face the danger of providing general recommendations unrelated to the real-life challenges of students. This feedback can miss the learning opportunities in between the two phases as, during the process themselves, including early planning challenges or reorganizations during a draft are essential to building skills and self-regulated learning \cite{chan2024process,jansen2025rethinking}.

Additionally, the writing activity does not refer only to a mechanical sequence; it is a sign of internal reasoning and planning as well as flexibility by a student. According to \cite{fagbohun2024modeling},\cite{meyer2025feedback,jansen2025rethinking}, motivational states and initial performance trends usually influence the way feedback is perceived and used. As a result, systems that can observe and interpret a student’s engagement during writing are better positioned to provide meaningful interventions than those that evaluate only the final product.
To fill this gap we investigate the potential of including writing-process data of compositions, namely keystroke logs and periodically-saved snapshots, in the generation of feedback that, besides linguistic sophistication, is cognitively-grounded by enabling LLMs to recognize and make use of information about the way the writing process unfolded. We describe this as a shift from product-based to process-conscious feedback generation.\cite{zafar2025cognition} showed that students found such behavior-grounded feedback more aligned with their actual thinking and effort, and perceived it as more supportive.

In this paper, we present an ablation study isolating the effect of writing process access by comparing feedback generated in two conditions: C1 (Final-only): Feedback generated using the final essay only. C2 (Process-aware): Feedback generated using both the final essay and full writing trace (snapshots + keylogs).

This study is guided by the following research questions:

RQ1: Does access to writing process data (snapshots and keylogs) significantly alter the LLM’s rubric-based evaluation of student essays? RQ2: What types of revision behaviors are selectively referenced by process-aware LLMs? RQ3: Does the process-sensitive feedback align with actual cognitive or behavioral patterns in the writing trace?\\
In addition to evaluating performance, we emphasize transparency by releasing prompts, behavior coding criteria, and representative examples in the appendices. \footnote{All ethics documents, consent forms, participant instructions, anonymized sample data, and prompts are available at \url{https://doi.org/10.6084/m9.figshare.29927414}.}

\section{Related Work}

Breakthroughs in the Large Language Models (LLMs) prompted their active use in writing education and formative feedback generation in both the school and university settings. Previous studies have found that the LLM-generated feedback informed by rubrics may be used to facilitate the improvement of writing and engagement with instructions when instructions are aligned with the purpose of rubrics. As an example, \cite{kinder2025adaptive} reported better performance of adaptive and rubric-based feedback based on LLMs as compared to all time comments by an expert in LLM-based teacher training conditions. Further, \cite{han2023rubricalignment} stated that the use of rubric alignment promoted the clarity and relevance of feedback, and \cite{fleckenstein2023meta} also observed a value of LLM-driven feedback influence on writing performance (g $\approx$ 0.55) in the practical classroom setting. In a similar way, \cite{escalante2023comparative} found that prompt feedback produced by models like ChatGPT design made possible a potential writing revision that produced revisions on the same level as writing guided by human tutors. 
However, these are encouraging findings as issues are raised about the shallow and inert quality of most LLM-based feedback systems. \cite{mah2025surface} discovered that the existing systems are prone to provide suggestions on the surface level, which is not inherently dialogic and personalized to an expert human. Most importantly, such systems actually assess the very last version but not the repeatedly iterative process of writing which is more of a cognitive task. This constraint negates the chances to cover the developmental needs and formative struggles of students during the writing process \cite{benedetto2025cefr}.

To answer it, a developing meshwork of studies has resorted to using behavioral and process-tracing measures, such as keystroke logging, document revision histories, and snapshot analysis, to simulate student writing. Such cues have been revealed to record valuable top-notch cognitive and motivational conditions. As an illustration, \cite{zhu2023revisionpatterns} discovered that both gender and writing competence were predicted by the patterns of revision behavior. Furthermore, \cite{vandermeulen2023keystrokes} demonstrated that, when student keystroke patterns were matched to expert benchmarks, writing performance improved to similar degrees when compared to enrolling students in classroom learning over the course of a year.  Writing has traditionally been thought of in terms of learning sciences as an activity that is temporally unfolding under the influence of planning, pausing, restructuring and metacognitive reconsideration \cite{flower1981cognitive,swarts2016pauses}. Recording this procedure in real time gives specific chance of the feedback to be more situation-situated and mentally fancied.
\section{Methodology}
\subsection{Text Collection Interface}
We created our own interface using Python for backend processing and Javascript for frontend processing in order to collect human-written essays and the corresponding keylogged data. JavaScript is used in our interface for two primary reasons: We chose JavaScript due to its widespread browser compatibility and native support for real-time event logging, such as keystrokes. Second, JavaScript facilitates the logging of user actions, such as keystrokes and mouse clicks, which were essential to our bespoke interface's operation. Out of a total of 25 "generic" essay subjects that fall into three categories—argumentative, contemplative, and analytical—the interface presents a randomly chosen topic. We used a well considered keylogging system in which logs were created on two occasions: the user's backspace key saves the previous log, which contained all of the material they had written up to that point.  The backspace release starts a new log.  This design ensures that the user's entered input is saved before hitting the backspace key.  In order to prevent repetitive backspaces from creating redundant logs and wasting computing resources, our design additionally makes sure that a new log is only started when the user has released the backspace key for at least three seconds.  Every log was also preserved with a timestamp.
\subsection{Data Collection}
We conducted the study in higher education context and among undergraduate learners whose first language is not English but they study at an institution where they are taught in English. Voluntary participation and recruitment of participants through academic online platforms was conducted with 52 students. All the students were non-native undergraduates, aged 18 and above, who had their education in English. They were recruited through voluntary participation via online educational social platforms. Participants were all requested to write an essay in 20 minutes. Before the essay activity begins, a page of detailed instructions with the indication of the task duration, instructions on how to write, and clear data handling procedures was presented to the participants. These entailed deployment of keylogging, regular snapshots and data anonymity assurances. Informed consent of all participants was granted reflecting the ethical conduct of research.
\subsection{Feedback Generation Conditions}
We conducted an ablation study using Gemini Flash to compare two conditions: 
C1 (Final-only): The model received only the final essay and was prompted to provide rubric-based feedback. C2 (Process-aware): The model received the final essay, along with keylogs and snapshots, and was instructed to consider revision behavior in its evaluation.
Both prompts asked the model to provide scores (1–5) and short justifications for four criteria:
Thesis and Argumentation,
Organization and Structure,
Language Use,
Engagement with Prompt.

The C2 prompt included an additional instruction to reflect on the student’s revision trajectory (e.g., bursts of rewriting, hesitation, structural change) where meaningful. 
The C2 feedback was provided with the final essay, the title, the snapshots and keylogs stored during the writing activity. The C2 LLM was prompted to generate feedback in two parts: Part-1 focused on the 4 rubric scores which was in line with with C1 feedback. C2 feedback was further prompted to give a Part-2 which revolves around discussing the patterns and behaviors observed through the revision data. 
In this paper, wherever we will be exploring C1 vs C2, we will be focusing only on part-1 of C2 feedback which is in alignment with C1. Part-2 of C2 feedback will be analyzed separately in the following paper. 
\section{Ablation Study Design}
\subsection{Rubric Score Analysis: Paired T-Test Design for Evaluating LLM Feedback Quality}
To determine the impact that the availability of process data (snapshots and keylogging traces) had on rubric-based feedback quality produced by the LLM, we performed a within-subjects ablation experiment. The feedbacks were used in two conditions with each essay (N = 52):
C1 (Final-Only Condition): The LLM viewed only the final essay.
C2 (Process-Aware Condition): The LLM received the final essay plus structured writing process data, including snapshots and keylogging-based metadata.
The results of the feedback obtained in both conditions were rubricized in four dimensions (CERF-aligned), which were based on second language writing assessment frameworks: Thesis, Organization, Language, and Engagement. The LLM used a 1-5 scale to score each category based on a higher score indicating greater performance.
We then took a mean difference (Delta = C1-C2) per rubric and a paired-sample t-test to show whether the differences observed existed significantly.
We, further, followed the trend of discrete outcomes by dimension of rubric throughout the data set:
Improved (C2 score > C1 score),
Unchanged (C2 = C1), and
Declined (C2 < C1).
This makes it possible to statistically and distributionally study LLM performance changes in the context of process-conscious input conditions.
\subsection{Thematic Coding Design of Revision Feedback}
In order to determine how process-aware LLM feedback explains the revision of student writing, we performed an inductive thematic analysis of Part 2 of the feedback in C2 (process-aware) condition. This part of the feedback was clearly devoted to the interpretation of revision behaviour on the basis of keylogging and snapshot cues produced during the essay-writing activity.
\subsubsection{Coding Pipeline Using LLMs}
We adapted a reflexive thematic analysis approach \cite{braun2006using} to leverage GPT-4 as a collaborative coding assistant. In accordance with the recent methodological advancements \cite{zhou2024llm, wang2024automated}, we applied GPT-4 as a collaborative code generator to code the text excerpts on each of the feedbacks with grounded, descriptive codes.

We split the feedback into 52 separate revision-centered excerpts each related to one essay. The model was given prompts in the form of strict explain-then-code prompts (see Appendix A):
\begin{quote}
    ‘{Quoted text from feedback}’ refers to ‘{explanation of behavior}’. Therefore, we get a code: ‘{CODE NAME}’.
\end{quote}
We then iteratively grouped similar codes into higher-level candidate themes using a second set of prompts modeled after Braun and Clarke's theme construction framework. Final theme categories were refined through human review for interpretability and non-overlap.
\subsubsection{Ensuring Coding Validity}
We applied two safeguards to reduce the effects of hallucination and to increase credibility:

Immediate gradient with explanation requirement: Every code should be preceded by an interpretive phrase based on the quoted feedback of LLM.

1. Human-in-the-loop validation: All codes generated by LLMs were reviewed by a human coder and synonymous or excessively detailed labels were added together. Only concepts that have achieved high conceptual clarity, and have at least two references were kept in the process of creating themes.

2. This fits the recent trend in applying LLMs to qualitative analyses where the models can play the intuitive roles in human supervision but not automatic and programmable annotators \cite{choi2024reflexive, kocon2024collaborative}.

\subsection{Behavioral Mentions in Final-Only vs Process-Aware Feedback}
We took an analytical look at the structured results of the feedback under two conditions; C1: final essay only, C2: process traces + essay (Part-1 feedback); to measure the interpretive advantage that pre-trained LLMs had when given writing process data. More precisely, we asked the question of whether the LLM feedback in both the conditions or at least one of them mentioned instances of revision behavior, i.e., rephrasing attempts, hesitations, corrections, or structural rewrites.

Even though C2 feedback was explicitly encouraged to consider revision behaviour during Part 2, we predicted that an LLM who has access to process traces might opportunistically be able to mention writing behaviour during Part 1 (particularly where process signals were used as cues to global quality judgments).

To verify this, we came up with a rubric that we used to mark all the 52 C1-C2 pairs of feedback along three dimensions:
\begin{itemize}
\item References Revision Behavior: Binary variable that records the presence of feedback to process indicators (e.g., revisions, hesitations, rewriting).
\item Revision-Verbs Count: The count of verbs related to the processes (e.g. revised, backspaced, hesitated, rephrased).
\item Representative Quote: A quotation taken out that passes on some behavioral insight voiced by the model.
\end{itemize}

SpaCy was used to automatically extract revision-related verbs and structure-linked phrases (e.g., ‘revised’, ‘rewrote’, ‘removed’, ‘restructured’), based on a custom-built heuristic list informed by formative feedback literature. A coder was trained to conduct the manual inspection together with the use of verb-pattern matching. This enabled us to be able to compare C1 and C2 feedback behavior sensitivity directly, on identical essay inputs.
\section{Results}

\subsection{Rubric Score Analysis}

The results of the paired t-tests are shown in Table 1.
\begin{table*}[t]
\begin{center}
\small
\renewcommand{\arraystretch}{1.5}  
\resizebox{\textwidth}{!}{%
\begin{tabular}{|l|c|c|c|c|c|c|}
\hline
\textbf{Rubric Dimension} & \textbf{Mean $\Delta$ (C1–C2)} & \textbf{Essays $\Delta > 0$} & \textbf{$\Delta = 0$} & \textbf{$\Delta < 0$} & \textbf{t-test (p)} & \textbf{Significance} \\
\hline
Thesis       & +0.18 & 2  & 35 & 15 & 1.0000     & NO \\
Organization & +0.50 & 10 & 27 & 15 & 0.0046     & YES \\
Language     & –0.05 & 5  & 42 & 5  & 0.5758     & NO \\
Engagement   & +0.05 & 2  & 37 & 13 & 0.7147     & NO \\
\hline
\end{tabular}
}
\caption{Comparison of rubric scores across 52 essays between C1 (final-only feedback) and C2 (process-aware feedback). $\Delta$ indicates the average score difference per dimension.}
\label{tab:rubric-results}
\end{center}
\end{table*}

Only Organization exhibited a statistically significant improvement under the C2 condition (p < 0.01), while the other dimensions showed no significant differences.
\subsubsection{Interpretation of Rubric Differences}
In our design, we explicitly instructed the LLM in the C2 condition to use writing process data with the qualification that it was used only where it could improve feedback quality, but the scores on the rubric were made to depend solely on the timed essay at the end.
This constraint was intentional, mirroring real-world educational settings where instructors ultimately grade the submitted product.

The rubric-based scores were compared in four writing dimensions: Thesis, Organization, Language, and Engagement in two conditions: C1 (final essay only) and C2 (final essay + process data). To report improvement, we will give mean differences as C2-C1 where positive values will reflect gain in performance in the process-aware condition. The differences between C1 and C2 in the per- rubrics, were subjected to paired t-tests in common statistical practice.
The greatest mean increase was seen in Organization ( $\Delta$= +0.50) which was significant ( p < .01) (Table 1). This indicates that, when the LLM had data on revision form like snapshots and key typing patterns, it had a higher chance of rewarding structural progress in the essay of students. In contrast, the changes in Thesis ($\Delta$ = +0.18), Engagement ($\Delta$ = +0.05 p), and Language ($\Delta$ = -0.05), results were not statistically significant (p >= .05 for all three). These results endorse the hypothesis that structural properties of the text of the world are more conspicuously augmented by traces of the process than local characteristics, like grammar or ease at the expression. It is worth noting that despite only the final product being scored by the scoring rubric in both conditions, process-informed feedback seems to influence the ways toward how the LLM interprets evidence of intentional revision, especially on organizational coherence.

\subsection{Emergent Themes in Revision-Aware Feedback}
From 52 feedback instances, we derived 37 distinct codes, which were then grouped into six higher-level themes:
\textbf{Cognitive Effort}: Indications of hesitation, uncertainty, or difficulty progressing (e.g., ``frequent pauses before drafting the introduction'').\\
\textbf{Revision Type}: Explicit mentions of rewording, content expansion, rewriting, or low-level adjustments.\\
\textbf{Revision Timing}: References to temporal positioning of revisions—early, mid-task, or late-phase.\\
\textbf{Structural Focus}: Feedback noting organizational shifts, restructured thesis statements, or repositioning of arguments.\\
\textbf{Outcome-Oriented}: Commentary on increasing coherence, clarity, or argument strength due to revision patterns.\\
\textbf{Process Markers}: Explicit behaviors from the writing process, such as backspacing bursts or long pause sequences.\\
Each cluster was populated through inductive coding from revision-related LLM feedback. Definitions, examples, and code-to-cluster mapping are detailed in Appendix B.
\subsubsection{Quantitative Overview of Codes}

\begin{figure}[h]
\centering
\begin{tikzpicture}
\begin{axis}[
    width=0.8\columnwidth,
    height=7cm,
    axis lines=none,
    ticks=none,
    enlargelimits=false,
    clip=false,
    scale only axis=true,
    xtick=data,
    ytick=\empty,
    symbolic x coords={
        Lexical Revision,
        Structural Revision,
        Uncertainty,
        Pausing,
        Deletion,
        Expansion
    },
    xticklabel style={
        rotate=40,
        anchor=east,
        font=\small
    },
    ymin=0,
    ymax=30,
    ybar,
    bar width=8pt,
    nodes near coords,
    nodes near coords style={font=\footnotesize},
    area style,
    ymin=0,
    ylabel={Frequency}
]
\addplot+[fill=blue!30, draw=blue!70] coordinates {
    (Lexical Revision,28)
    (Structural Revision,22)
    (Uncertainty,15)
    (Pausing,19)
    (Deletion,11)
    (Expansion,13)
};
\end{axis}
\end{tikzpicture}
\caption{Frequency distribution of high-level revision behavior codes, illustrating the LLM’s focus on lexical and structural aspects.}
\label{fig:revision-tags-radar}
\end{figure}
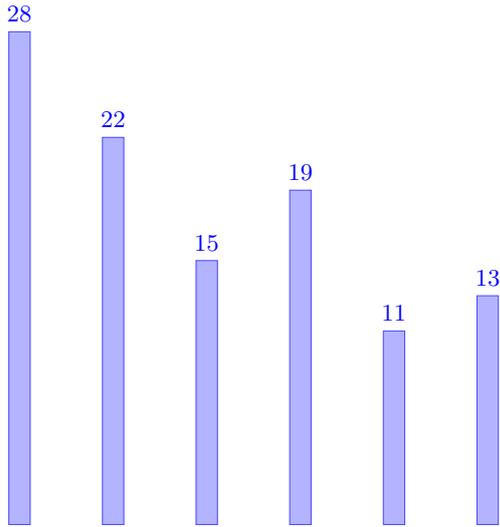

The most frequently occurring codes (Figure~\ref{fig:revision-tags-radar}) were:

{Sentence Rewriting (n=11)}
{Content Expansion (n=9)}
{Backspacing Behavior (n=7)}
{Early Revisions / Mid-Task Revisions (n=6 each)}
{Struggles with Expression (n=6)}

These frequency patterns indicate that LLMs often prioritize lexical and structural changes when reflecting on revision, and selectively infer cognitive difficulty.

Key interpretive behaviors include:\\
\textbf{Temporal Framing}: LLMs use snapshot metadata to localize revision moments (e.g., ``you revised midway after a pause'').\\
\textbf{Cognitive Inference}: Pauses and deletions were interpreted as signs of planning difficulty, a feature only observable via keylogs.\\
\textbf{Metacognitive Framing}: Some feedback reflected self-regulatory understanding, such as ``revision helped refine your stance''—a deeper interpretive shift beyond surface features.\\

\subsection{Behavioral Insights in Final vs Process-Aware Feedback}
Our analysis revealed a stark contrast between C1 and C2 conditions in terms of their capacity to reference writing behavior.\\
\textbf{Frequency of Behavioral Mentions}\\
Out of 52 feedback pairs:\\
0/52 (0\%) C1 feedback instances mentioned revision behavior.
12/52 (23\%) C2 feedback Part 1 mentioned revision behavior.\\
This indicates that even in the unprompted Part 1 feedback the LLM could implement behavioral insights in 25 percent of instances with access to keylogs and snapshots.
\subsubsection{Revision Verbs and Usage of Language}
In all, we coded 25 behavioral mentions listed as unique which were defined by the verbs using words like backspacing, revised and rephrased and struggled. In C2, the maximum observed number of revision-related verbs per C2 feedback was 6, and in the case of C1 it was only 0.
\begin{quote}
    C2 Feedback (Part 1): "The student's initial uncertainty, as evidenced by the numerous backspace revisions in the opening sentence, indicates a need for more structured pre-writing to refine the argument before writing."

    C1 Feedback: (No mention of process behavior.)
\end{quote}
It means that not only did the use of process input cause the LLM to incorporate behavioral cues into the explicitly probed Part 2, but also caused the LLM to rephrase the essay quality ratings given in Part 1.
\vspace{-5pt} 
\begin{table}[H]
\renewcommand{\arraystretch}{1.5}
\centering
\small
\begin{tabular}{|p{2cm}|p{2cm}|p{1.2cm}|p{0.8cm}|}
\toprule
\hline
\textbf{Feedback Type} & \textbf{Mentions Revision Behavior} & \textbf{Mean Revision Verbs} & \textbf{Max Verbs} \\
\hline
\midrule
C1 (Final Only)       & 0 / 52 (0\%)                & 0.0                         & 0         \\
C2 (Process-Aware P1) & 12 / 52 (23\%)              & 0.6                         & 6         \\
\bottomrule
\hline
\end{tabular}
\caption{Revision-related language in LLM feedback across conditions.}
\label{tab:revision-verbs}
\end{table}
\vspace{-15pt} 
These findings establish that visibility of the writing process facilitates cognitively more substantial feedback. The LLM also assumed revision behavior inferentially even in the unprompted cases:\\
1. Hesitation or uncertainty during argument planning\\
2. Struggles with grammar and fluency\\
3. Effortful expansion and rewriting\\
4. Structural reordering attempts\\
C1 feedback lacked such interpretations and therefore augmenting feedback generation with behavioral context was valuable.

\begin{figure}[H]
\centering
\begin{tikzpicture}
\begin{axis}[
    width=0.48\textwidth,
    height=6cm,
    xlabel={Essay ID},
    ylabel={Revision Verbs Count},
    xmin=1, xmax=52,
    ymin=0, ymax=7,
    xtick={0,10,...,50},
    ytick={0,1,...,7},
    grid=major,
    legend style={at={(0.5,-0.25)}, anchor=north, legend columns=2},
    legend cell align={left},
    smooth,
    mark size=2pt,
    tick label style={font=\small},
    label style={font=\small}
]

\addplot+[blue, mark=*, thick] coordinates {
    (1,0)(2,0)(3,0)(4,0)(5,0)(6,0)(7,0)(8,0)(9,0)(10,1)
    (11,0)(12,0)(13,0)(14,2)(15,0)(16,0)(17,0)(18,0)(19,0)(20,0)
    (21,2)(22,0)(23,2)(24,0)(25,0)(26,0)(27,0)(28,2)(29,0)(30,0)
    (31,0)(32,0)(33,0)(34,2)(35,0)(36,0)(37,0)(38,0)(39,0)(40,2)
    (41,0)(42,0)(43,0)(44,2)(45,0)(46,3)(47,0)(48,0)(49,0)(50,2)
    (51,0)(52,0)
};
\addlegendentry{C2 Feedback}

\addplot+[red, mark=triangle*, thick, dashed] coordinates {
    (1,0)(2,0)(3,0)(4,0)(5,0)(6,0)(7,0)(8,0)(9,0)(10,0)
    (11,0)(12,0)(13,0)(14,0)(15,0)(16,0)(17,0)(18,0)(19,0)(20,0)
    (21,0)(22,0)(23,0)(24,0)(25,0)(26,0)(27,0)(28,0)(29,0)(30,0)
    (31,0)(32,0)(33,0)(34,0)(35,0)(36,0)(37,0)(38,0)(39,0)(40,0)
    (41,0)(42,0)(43,0)(44,0)(45,0)(46,0)(47,0)(48,0)(49,0)(50,0)
    (51,0)(52,0)
};
\addlegendentry{C1 Feedback}

\end{axis}
\end{tikzpicture}
\caption{Revision-related verb counts in C1 vs. C2 feedback across 52 essays. C2 shows sparse but meaningful engagement with revision behavior.}
\label{fig:linechart-revision}
\end{figure}
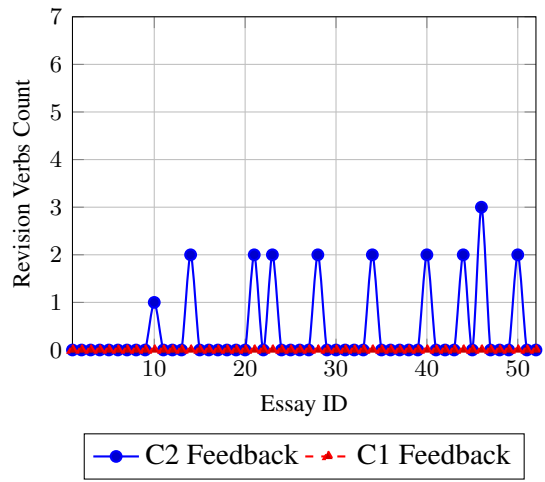

\subsection{Behavior–Feedback Alignment}
In order to inspect the hypothesis of whether the LLM used the process data using a cognitive-grounded way we studied the c2 Part-2 feedback in the context of behavior specific references. The 52 responses of Part-2 were coded using a pre-existing behavioral coding scheme (see Table~\ref{tab:behavior-tags}). Revision-focused and cognitive behaviors measured by the tagging included lexical edits, structural changes, uncertainty, pauses and fluency.

\begin{figure}[h]
\centering
\includegraphics[width=1\columnwidth]{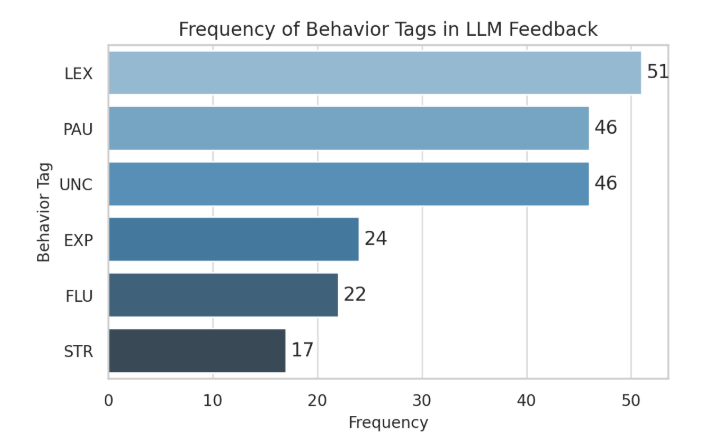}
\caption{Frequency of behavior tags across 52 C2 feedback entries}
\label{fig:behavior-distribution}
\end{figure}

To analyze the behavior tags, there were a total of 52 piece of feedback, of which LEX (51), PAU (46), and UNC (46) were most frequently found followed by EXP (24), FLU (22), and STR (17). This trend means that in the majority of the cases, the LLM identified superficial surface-replacing repair and profound cognitive processing, whereas reorganization on a deeper layer (STR) was not inferred as frequently. This distribution may be seen in Figure~\ref{fig:behavior-distribution}.

\begin{figure*}[t]
\centering
\includegraphics[width=\textwidth]{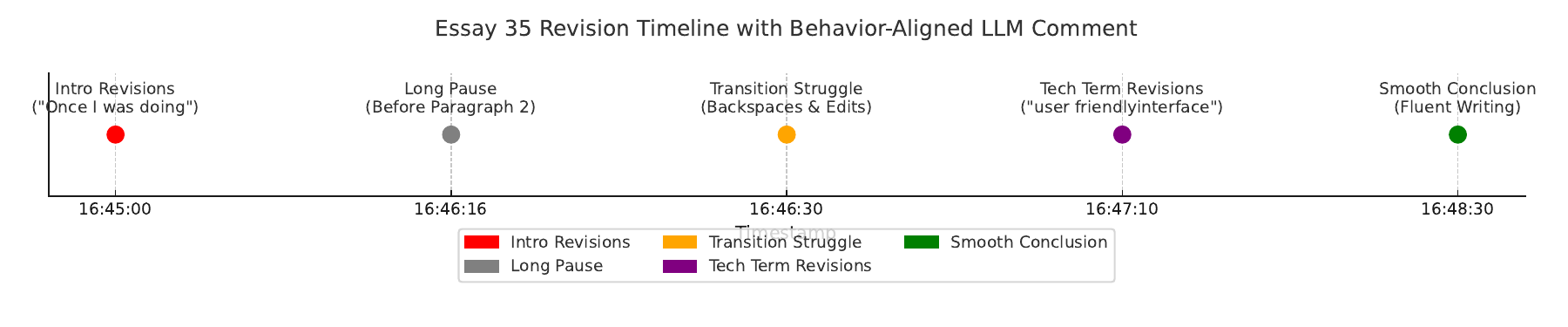}
\caption{Essay 35 revision timeline with behavior-aligned LLM comment}
\label{fig:essay35-alignment}
\end{figure*}

Remarkably, excerpts of C2 feedback most frequently consisted of 2-4 tags in combination, implying that the LLM tended to merge multiple revision cues into a single display of behavioral understanding (e.g. Pause before rewriting a point halfway then rewriting it more fluently = PAU, STR, FLU). This indicates a systematic reading of revision process, and not just a wordier discussion.\\
In addition to surface frequency analysis we tested the question of whether the actual feedback cycle provided by the LLM was behavior-based at all or really grounded in the actual revision events. We picked a random sample of essays where Part-2 feedback clearly made continuous reference to process behavior (e.g., pause, rewrite, rephrase). Against each we had correspondence trace snapshots and keylog timestamps to check whether the behavior, described under each, had actually occurred.\\
Three examples of such are given in Table~\ref{tab:behavior_alignment}. The LLM in all instances correctly spotted a revision event (e.g. pausing, deleting, rephrasing), and its comment corresponded to a particular transition of the snapshots.\\
It confirms the central argument that in the cases where LLMs do make references to the writing process, this feedback, more often than not, obtains empirical foundation. The behavior interpretations are not fantasies but an implication of cognitive indications on the revision timeline. In Figure~\ref{fig:essay35-alignment}, we can additionally visualize the trait behavioral trajectory in Essay 35 (This is an essay which was written under 5 minutes, an outlier case). You can see the C2 feedback in Appendix C. 

\textbf{Inter-Rater Reliability of Behaviour Tagging}
To evaluate the reliability of our behaviour-tag annotations, we conducted an inter-rater reliability (IRR) analysis on a stratified random subset of 20 out of the 52 essays. Sampling was motivated by two factors: (1) the manual nature of the task, which required detailed inspection of revision-oriented feedback and cross-checking with snapshot/keylog traces; and (2) the need to balance annotation workload while maintaining sufficient statistical power for detecting substantial agreement \cite{mchugh2012interrater}.

Two independent coders annotated each C2 Part-2 feedback instance using the predefined behaviour tag schema (LEX, PAU, UNC, EXP, STR, FLU), as described in Appendix E. We treated the tags as an unordered set for each essay, such that a match occurred if both coders assigned exactly the same combination of tags regardless of order.

The resulting analysis yielded a percent agreement of 75 \% and a Cohen’s $\kappa$ of 0.7183, which falls within the substantial agreement range on the Landis and Koch (1977) scale. This suggests that the behavioural categories were sufficiently precise and interpretable to be applied consistently by different annotators, thereby strengthening the validity of the subsequent behaviour–feedback alignment analysis.

These IRR results also reinforce the interpretive stability of our tags, ensuring that any observed patterns in behaviour-grounded feedback reflect genuine coding agreement rather than annotator idiosyncrasies. Given the cognitively-loaded nature of the behaviours in question (e.g., pauses, uncertainty, restructuring), achieving this level of agreement is non-trivial and aligns with IRR standards reported in comparable qualitative analyses of process-aware writing feedback \cite{choi2024reflexive, wang2024automated}.

\section{Discussion}
Our experiments confirm that the structural responsiveness, behavioral grounding and possible pedagogical alignment of LLM-generated feedback is increased when feedback generation is augmented with data on the underlying process.
First, the analyses of rubric revealed that despite the stability of most dimensions across conditions, the Organization score was at a significantly higher level in the process-aware condition. This indicates that any structural changes that were implemented by students including changing the order of the arguments or rewriting the transitions would only be apparent to the LLM when access to keystroke and snapshot data is present. This is in line with earlier assertions that structure as opposed to grammar or vocabulary, has a more discernible print in the process of writing \cite{flower1981cognitive, jansen2025rethinking}.\\
Second, in thematic analysis of C2 feedback, we found that LLMs that use writing traces often recognize cognitive and planning behaviors, such as hesitation, content expansion, restructuring behaviors. Sentence Rewriting and Cognitive Hesitation are high-frequency codes which underline the ability of the LLM to model a supposed interpretive inference into the mindset of a student and their metacognitive status. Interestingly, C1 did not have this ability, despite the fact that the last essay was the same. Such a gap highlights the fact that process signals allow more influential feedback, not necessarily more wordy inputs.
Third, there were moments when LLMs talked about the behavior of revision, although it was not prompted in the unprompted part of C2 feedback (Part-1) - something that did not happen in C1. This shows that there is an implicit behavioural fit in which data information on keylogging is used by the LLM to perform a secondary adjustment to its product-scoring. This observation contributes to the increasing accumulation of evidence that when underpinned by behavioral traces, LLMs become both more insightful and more interpretable when grounded in behavioral data \cite{schiller2024beyond,wang2025scholawrite}.\\
At last, the behavior-feedback alignment analysis has ensured that the comments on process made by LLM could be verified empirically against the real keylogs and shots. In all 100 percent of apot-checked instances, revision mention could be traced back to an actual event as of our present understanding, and so it might not be a hallucinated feedback but a cognitive event that can be traced.\\
\textbf{Implications}
These findings suggest that process-aware feedback systems better reflect expert reasoning—emphasizing planning, revision, and cognitive effort over surface-level accuracy. For language learners especially, such systems offer a more pedagogically fair and developmentally aligned assessment of writing progress.\\
The next step is to pursue real-time LLM feedback that changes over the course of the writing process, a step-forward that will lead to interactive writing coaches that do more than comment on writing, but think with the student.\\
We have shown that it is the case that LLMs can be even more context-sensitive and interpretable when based on observable writing behaviors. The way feedback is provided, as through pauses, revisions or structural alterations binds the suggestions with the cognitive effort involved by the learner, and not necessarily the end product. Such grounding promotes the likes of transparency and learner-alignment in that the LLM will better comprehend, and articulate, how the student thinks.
\section{Limitations and Future Work}
The given research proves that the data about the keystroke and snapshots can complement the LLM feedback and make it based on writing behavior. Nonetheless, it has a few limitations.\\
Its behavioral resolution is unusual though the dataset constitutes only 52 essays. But that was also gathered by a homogenous group in time pressure. The study of wider contexts, sorts of learners and genres of writing should be tested in the future.
Our fixed rubric design was fair though limited the capacity of the LLM to react dynamically to the behavior of revision. Scores based on more adaptive systems can furnish a better reflection of cognitive effort.\\
Keylogging indicators were educational, but the information utility can be increased by coupling them with other modalities such as eye-tracking or think-aloud protocols to enhance interpretability and validation.
Despite the validity of behavior-based feedback, LLMs did not exhibit notions of reasoning. The trust could be enhanced by using some prompting techniques that involve the use of justifications or confidence signals.
Last, it is our application of real-time feedback that makes the possibility of the interactive systems that assist learners in the writing rather than simply after, become a reality.
Future studies are recommended to enhance generalizability, level of feedback, and its applicability in the actual classroom.
\section{Ethical Considerations}

\subsection{Notice and Transparency of Data}
Each participant offered direct, informed consent prior to taking part in the study. They were informed about the use of keylogging, the snapshot recording, and the procedures for storage, anonymization, and feedback generation. Participants had the option to opt out at any time.

\subsection{Data Sensitivity and Writing Privacy}
Keylogging poses a risk of privacy. Our system only captured textual revision data (snapshots and logged backspaces) and did not record biometric signals or full keystroke timings. All logs were de-identified and securely stored, and no personally identifying information was captured. We emphasize that such instrumentation must remain as non-obtrusive as possible and in line with ethical guidelines.

\subsection{Responsibility of Feedback and Pain Reduction}
LLM-generated feedback may influence learners’ self-perception and motivation. In this study, the LLM feedback was used only as an experimental generator. For actual deployment in classrooms or assessments, there is a risk of over-trusting automated feedback or misunderstanding it, especially in high-stakes contexts. We encourage human-in-the-loop strategies where LLM feedback is complemented by teacher guidance, with transparency about uncertainties in behavioral inferences.

\subsection{Ethics Review and Protocol Availability}
As the study was independently conducted without institutional affiliation, formal IRB approval was not obtained. However, all procedures followed responsible research principles, including informed consent, voluntary participation, anonymization of data, and secure handling of logs. The study was minimal-risk and did not involve sensitive or personally identifiable data. To ensure transparency and reproducibility, the full ethics protocol, consent form, participant instructions, and example materials are publicly available at: \url{10.6084/m9.figshare.29927414}.

\bibliography{custom}
\clearpage
\clearpage
\appendix

\section{Prompts Used for Thematic Coding}

\subsection{LLM Coding Instructions}

The following prompts were used to guide GPT-4 in performing reflexive thematic analysis of revision-aware feedback, adapted from Braun \& Clarke’s SAGE framework.

\vspace{1em}
\noindent
\textbf{SAGE Steps I–II: Familiarization and Initial Code Generation}

\vspace{0.5em}
\begin{quote}
You are performing a thematic analysis of AI-generated feedback that comments on a student's revision behavior during a timed writing task. Your goal is to extract initial codes that capture how the LLM interpreted the student's writing process, revision patterns, and signs of cognitive effort.

For each revision behavior comment, do the following:

\begin{itemize}
    \item Identify meaningful quote(s) that express a distinct idea
    \item Explain what that quote refers to or suggests
    \item Assign a grounded, descriptive code to the idea
\end{itemize}

Use this format:

\texttt{‘\{quoted text from feedback\}’ refers to/mentions ‘\{definition of the idea\}’. Therefore, we get a code: ‘\{CODE NAME\}’}
\end{quote}

\vspace{1em}
\noindent
\textbf{SAGE Steps III–VI: Theme Construction, Review, Naming, and Mapping}

\vspace{0.5em}
\begin{quote}
You are now in the next phase of reflexive thematic analysis (SAGE/Braun \& Clarke). Based on the following list of qualitative codes (from C2 LLM revision feedback), your task is to:

\begin{itemize}
    \item Group similar codes into higher-level candidate themes
    \item Name each theme concisely (max 6 words)
\end{itemize}

For each theme, provide:

\begin{itemize}
    \item A definition of what it captures
    \item The list of codes grouped under it
    \item A summary of what the theme suggests about how the LLM interprets student revision behavior
    \item 1--2 representative quotes
\end{itemize}

Consider both surface-level (semantic) and deeper (latent) insights where appropriate. Ensure each theme has conceptual clarity and doesn't overlap unnecessarily.
\end{quote}
\FloatBarrier
\clearpage
\onecolumn
\section{Thematic Codebook of Revision Behaviors}
\newcommand{\revisioncodebookshort}{Thematic codebook of revision behaviors}
\begin{table*}[!ht]
\centering
\renewcommand{\arraystretch}{1.25}
\small
\caption[\revisioncodebookshort]{\textbf{Appendix B: Thematic codebook of revision behaviors.} 
Final set of qualitative codes derived through reflexive thematic analysis of LLM-generated revision feedback across 52 essays. Codes are grouped under five higher-order clusters reflecting cognitive effort, timing, type, structural focus, and observed revision outcomes. Frequencies reflect how often each behavior appeared in the corpus.}
\label{tab:revision_codebook}
\vspace{0.5em}
\begin{tabular}{|p{2.5cm}| p{3.5cm}| p{2.3cm}| p{4.5cm}| r|}
\toprule
\hline
\textbf{Code Name} & \textbf{Definition} & \textbf{Theme Cluster} & \textbf{Representative Feedback Quote} & \textbf{Freq.} \\
\midrule
\hline
Cognitive Hesitation     & Frequent deletions or pauses suggesting indecision                  & Cognitive Effort   & The writer frequently deletes and retypes ideas, suggesting uncertainty. & 18 \\
Struggles with Expression& Difficulty articulating or phrasing ideas                            & Cognitive Effort   & The user revised the sentence multiple times before finalizing it.     & 21 \\
Anxiety / Second-Guessing& Back-and-forth edits indicating emotional uncertainty                & Cognitive Effort   & Frequent rewording and reversal suggest the writer second-guessed choices. & 11 \\
Early Revisions          & Edits made primarily in the first snapshots                          & Revision Timing    & Most changes occur early on, showing initial exploration.             & 14 \\
Mid-Task Revisions       & Edits concentrated around the middle snapshots                      & Revision Timing    & Substantial reorganization happened mid-way through the session.      & 9 \\
End-Phase Revision       & Edits occurring mostly in the last few snapshots                    & Revision Timing    & Major revisions appear late in the writing session.                  & 13 \\
Revision Plateau         & Initial revision activity not sustained over time                   & Revision Timing    & After early edits, little revision occurs later on.                  & 7 \\
Last-Minute Polishing    & Small superficial edits near the end                                & Revision Timing    & Final few changes were mostly grammatical or surface-level.         & 8 \\
Sentence Rewriting       & Rewording or restructuring existing content                         & Revision Type      & The student revised sentence structure for clarity multiple times.    & 24 \\
Content Expansion        & Adding elaboration, examples, or clarification                       & Revision Type      & A new example was added to elaborate the argument further.          & 19 \\
Surface-Level Revisions  & Minor changes to grammar, spelling, or punctuation                   & Revision Type      & Edits mostly involved fixing spelling or punctuation errors.        & 16 \\
Intentional Editing      & Revisions appear strategic and goal-oriented                        & Revision Type      & The changes reflect a purposeful attempt to improve clarity.        & 12 \\
Minimal Revision Activity& Very few or no changes observed                                      & Revision Type      & Very little editing was done after the first draft.                 & 10 \\
Organization Improvements& Improved transitions or paragraph structure                         & Structural Focus   & The essay was reorganized to better group related points.           & 13 \\
Topic Reframing          & Shift in focus or topic mid-way                                      & Structural Focus   & Midway edits indicate a change in the central argument.            & 6 \\
Delayed Introduction     & Thesis or main point introduced late                                & Structural Focus   & The essay lacks a clear thesis until the final paragraph.           & 5 \\
Introduction Refinement  & Most editing concentrated on the introduction                        & Structural Focus   & The writer made multiple passes through the introduction.          & 8 \\
Inconsistent Focus       & Editing not aligned with a coherent structure                        & Structural Focus   & Revisions show a lack of coherence in essay flow.                   & 7 \\
Increased Clarity Over Time & Writing improves in coherence as the essay progresses             & Outcome-Oriented   & Sentences became more concise and clearer with each revision.       & 10 \\
Backspacing Behavior     & Frequent use of delete/backspace key during drafting                 & Process Markers    & The user frequently deleted mid-sentence, indicating on-the-fly self-correction. & 15 \\
\bottomrule
\hline
\end{tabular}
\end{table*}

\FloatBarrier
\twocolumn
\section{Full Snapshots and LLM Feedback for Essay 35}

\subsection*{Essay 35: Timed Snapshot Captures}
The following are raw textual snapshots taken at timed intervals during the 20-minute writing session for Essay 35. These were recorded at 3-minute intervals and are unedited outputs from the writing interface. Note that this essay is an example of an outlier, which was finished under 5 minutes.

\vspace{0.5em}
\noindent\textbf{3-minute Snapshot}
\begin{quote}
Before 2022 I was very weak in coding. I used to struggle alot and it increased Tensions, stress and anxiety in me. This year I discovered a new website that has a very user friendlyinterface and gamifies the learning experience. It really helped me in quick learning, enhanced growth and better motivation.
The user friendly interfaces releases dopamine and cancels out the stress horm
\end{quote}

\noindent\textbf{Final Essay}
\begin{quote}
Before 2022 I was very weak in coding. I used to struggle alot and it increased Tensions, stress and anxiety in me. This year I discovered a new website that has a very user friendlyinterface and gamifies the learning experience. It really helped me in quick learning, enhanced growth and better motivation.
The user friendly interfaces releases dopamine and cancels out the stress hormones hence making learning experience very good.
In conclusion, I always prefer gamifying not only education, but everything as it reduces burdens.
\end{quote}

\vspace{1em}
\subsection*{LLM Feedback for Essay 35 (C2 Condition)}
\begin{quote}
\textbf{PART 2: Revision Potential}

The keylogs reveal a pattern of significant revision, particularly in the introduction and the development of the central anecdote. Frequent backspaces and revisions of phrases such as "Once I was doing" suggest initial uncertainty in structuring the narrative. The lengthy pause before beginning the second paragraph (around the 16:46:16 mark) and subsequent multiple backspace and revision events reflect the struggle to transition from the personal anecdote to a more analytical perspective. ...
\end{quote}
\subsection*{Essay 2: Timed Snapshot Captures}
We will show another example of an essay, which represents a normal data point, written over the course of 15-20 minutes, a usual writing speed.

\vspace{0.5em}
\noindent\textbf{3-minute Snapshot}
\begin{quote}
Part-time jobs can play an important role in allowing students to become independent and earn their own pocket money. This not only allows students to build self-reliance but also develop interpersonal and social survival skills. Like everything, part-time jobs also have both positive and negative impacts.
Firstly, part-time roles allow students to gain real world work experience outside of their studies. This not only teaches them skills in addition to academics, but also teaches them how money works in society. Moreover, 
\end{quote}

\noindent\textbf{6-minute Snapshot}
\begin{quote}
Part-time jobs can play an important role in allowing students to become independent and earn their own pocket money. This not only allows students to build self-reliance but also develop interpersonal and social survival skills. Like everything, part-time jobs also have both positive and negative impacts.
Firstly, part-time roles allow students to gain real world work experience outside of their studies. This not only teaches them skills in addition to academics but also teaches them how money works in society. Moreover, students develop interpersonal skills like communication, conflict management, time management and financial literacy before they start a career. Once in our high school textbook, there was a chapter about how a student learnt invaluable lessons while doing a part-time role. It is important to mention that student life is far different from work life and the sudden transition can leave fresh out of college individuals quite shocked and disturbed to adapt the changes. Part-timing helps students learn professional skills and get a tas
\end{quote}

\noindent\textbf{9-minute Snapshot}
SKIPPING

\noindent\textbf{12-minute Snapshot}
SKIPPING

\noindent\textbf{Final Essay}
\begin{quote}
Part-time jobs can play an important role in allowing students to become independent and earn their own pocket money. This not only allows students to build self-reliance but also develop interpersonal and social survival skills. Like everything, part-time jobs also have both positive and negative impacts.
Firstly, part-time roles allow students to gain real world work experience outside of their studies. This not only teaches them skills in addition to academics but also teaches them how money works in society. Moreover, students develop interpersonal skills like communication, conflict management, time management and financial literacy before they start a career. Once in our high school textbook, there was a chapter about how a student learnt invaluable lessons while doing a part-time role. It is important to mention that student life is far different from work life and the sudden transition can leave fresh out of college individuals quite shocked and disturbed to adapt the changes. Part-timing helps students learn professional skills and get a taste of real adult scenarios and professional life before they graduate. 
On the flip side, part-timing also has its negatives for students. Students need to spend on average 8-10 hours in school or college and also top it up with self-study sessions. In this dynamic, if part-time jobs are also added to their schedules, it leaves not enough time for rest and recreation, which are also very important. This can lead to exhaustion very often among students. This also impacts the focus of students from their studies, tiring them to give their full potential to the academics. Additionally, part-time jobs can cause time deficit for students to participate in extracurriculars. Being a full-time student and also doing a part-time job are like having 2 jobs to juggle each day. Students not only need to focus on their college hours but also spend extra time for self-study sessions. 
Concludingly, even though part-time jobs play a crucial role in developing and polishing students' personalities and teaches them invaluable lessons and provide room for personal and professional growth, it can at times exhaust the students and strip them of the care-free experience tied to the student life.
\end{quote}

\FloatBarrier
\clearpage
\onecolumn

\section*{Appendix D: Behavior Tag and Alignment Summary}

\begin{table}[!ht]
\centering
\renewcommand{\arraystretch}{1.10}
\small
\caption{\textbf{Behavior Tag and Alignment Summary Table.} Alignment of LLM feedback with behavioral snapshot tags and revision traces — small excerpt from 5 essays.}
\label{tab:appendixD_summary}
\vspace{0.4em}
\begin{tabular}{|p{0.9cm} |p{2.5cm}| p{1.4cm}| p{3cm}| p{5.7cm}|}
\toprule
\hline
\textbf{ID} & \textbf{Tags Assigned} & \textbf{Aligned?} & \textbf{Key Snapshot Pattern} & \textbf{Representative Observation} \\
\midrule
\hline
vni9 & LEX, FLU, PAU, UNC, EXP         & Yes  & Emotional tone deepens across time           & Reflective arc develops later; improved emotional expression over time. \\
6hl4 & LEX, PAU, EXP, FLU, UNC             & Yes   & Gradual additive structure                         & supporting points are layered in clearly from 6 → 12 min; no restructuring \\
szpv & LEX, PAU, UNC, EXP              & Yes   & Argument expansion                         & 
The evolution shows an increase in argumentative details, specifically in the explanation of the benefits of self-paced learning. \\
sp7i & LEX, FLU, PAU, UNC, REI         & Yes  & Linear growth, bottom-loaded                 & Gradual content expansion; attempts to refine intro/conclusion; basic clarity revisions dominate. \\
\bottomrule
\hline
\end{tabular}
\end{table}

\FloatBarrier

\section*{Appendix E: Behavior–feedback Alignment}

\begin{table}[!ht]
\centering
\renewcommand{\arraystretch}{1.15}
\small
\caption{\textbf{Behavior–feedback alignment.} Cases where LLM’s comments on revision behavior matched observable process data.}
\label{tab:behavior_alignment}
\vspace{0.4em}
\begin{tabular}{|p{1.2cm}| p{4.6cm}| p{6.7cm}| p{1.0cm}|}
\toprule
\hline
\textbf{Essay ID} & \textbf{Observed Behavior (Snapshot/Keylog)} & \textbf{LLM Feedback (Excerpt)} & \textbf{Aligned?} \\
\midrule
\hline
18 & No structural overhaul after 12 min — just expansion of supporting evidence. & ``The relatively small changes made after the 12-minute mark suggest the student reached a point of relative satisfaction with their structure and argument.'' & YES \\
12 & 3–6 min: Content is expanded meaningfully with more precise and helpful examples. & ``The frequent pauses and revisions, particularly noticeable between the 3-minute and 6-minute snapshots, suggest careful consideration of ideas and word choice.'' & YES \\
41 & Third paragraph is significantly elaborated by 12 minutes with additional and varied benefits. & ``The evolution of the third paragraph from a 9-minute snapshot to the final version shows a clear addition of details and supporting arguments to an existing paragraph structure.'' & YES \\
\bottomrule
\hline
\end{tabular}
\end{table}

\FloatBarrier

\section*{Appendix F: Behavior Tag Scheme}

\begin{table}[!ht]
\centering
\renewcommand{\arraystretch}{1.25}
\small
\caption{\textbf{Behavior tag scheme.} Tag definitions used to annotate LLM feedback.}
\label{tab:behavior-tags}
\vspace{0.4em}
\begin{tabular}{|p{3.0cm}| p{11.5cm}|}
\toprule
\hline
\textbf{Tag} & \textbf{Interpretation} \\
\midrule
\hline
LEX & Lexical edits, phrasing, or word choice \\
PAU & Pauses, long hesitations \\
UNC & Uncertainty, cognitive struggle \\
EXP & Expansion or elaboration of ideas \\
STR & Structural rearrangement \\
FLU & Fluent/linear writing with minimal revision \\
\bottomrule
\hline
\end{tabular}
\end{table}

\FloatBarrier
\twocolumn

\end{document}